\documentclass[12pt]{iopart}
\usepackage[T1]{fontenc}
\usepackage[latin9]{inputenc}
\usepackage{amssymb}
\usepackage{graphicx}



\usepackage[english]{babel}
\begin{document}

\title{The Elliptic Model for communication fluxes}

\author{C Herrera-Yagüe$^{1,2,3}$, CM Schneider$^1$, Z Smoreda$^4$, T Couronné$^4$, PJ Zufiria$^{2,3}$ and MC González$^1$}

\address{$^1$Department of Civil and Environmental Engineering, Massachusetts Institute of Technology, Cambridge, Massachusetts, United States of America}
\address{$^2$Depto. Matemática Aplicada a las Tecnologías de la Información, ETSI Telecomunicación, Universidad Politécnica de Madrid (UPM), Spain}
\address{$^3$Cátedra Orange. Universidad Politécnica de Madrid (UPM), Spain}
\address{$^4$Sociology and Economics of Networks and Services department, Orange Labs, Issy les Moulineaux, France}

\eads{\mailto{carlos@hyague.es}}
\begin{abstract}
    In this paper, a model (called Elliptic Model) is proposed to estimate the 
    number of social ties between two locations using population data in a 
    similar manner transportation research does with trips. To overcome the 
    asymmetry of transportation models, the new model considers that the number 
    of relationships between two locations is reversely proportional to the 
    population in the ellipse whose focuses are in these two locations. The 
    Elliptic Model is evaluated by considering the anonymous communications 
    patterns of 25 million users from 3 different countries, where a location 
    has been assigned to each user based on her most used phone tower or 
    billing zip code. With this information, spatial social networks are built 
    at three levels of resolution: tower, city and region for each of the three 
    countries. The Elliptic Model reaches similar performance when predicting 
    communication fluxes as transportation models do when predicting trips.  
    This shows that human relationships are at least as influenced by geography 
    as human mobility is.  \end{abstract}
\maketitle

\section{Introduction}

While social networks have been known for years to play a key role
in various human phenomena \cite{granovetter1973strength,milgram1967small},
 during decades their study was limited to certain kind of social relationships 
 for which interaction records were available, such as authorship and 
 cooperation in science \cite{jones2008multi,cairncross2001death}. Only 
 recently it has been possible to map large social networks representing a 
 broader range of interactions in order to explore how their structures 
 influence processes occurring in these networks.  The required large social 
 network data sets, usually coming from telecommunication records originated in 
 e-mail \cite{kossinets2006empirical},
phone \cite{onnela2007structure} or online communication platforms
\cite{mislove2007measurement}, have been used to explore a wide range
of topics such as adoption of innovation \cite{toole2012modeling},
social groups discovery \cite{yang2012community,ahn2010link,blondel2008fast},
epidemic spreading \cite{pastor2001epidemic,wang2009understanding,schneider2011suppressing}, social mobilization
\cite{pickard2011time} or sentiment spreading \cite{fowler2008dynamic}. 

Despite the publication of such studies, network data is not widely
available to the community due to legal, privacy or commercial issues.
In addition, even with access to the electronic records, extracting
a meaningful social network may be difficult at a large
scale \cite{onnela2007structure}. For these reasons, creating
models that are able to mimic different social network properties
have recently attracted a fair amount of research interest \cite{watts1998collective,barabasi1999emergence,karrer2010random,holme2002growing,herrera2011generating}.
While most models try to generate synthetic networks with some desired
characteristics (degree distribution and clustering coefficient among others),
we will focus here on reproducing a macroscopic feature of
real social networks: the number of social ties between different locations,
i.e. how many relationships exist between two cities, two regions
or even two neighborhoods. Throughout the paper, we will employ the term 
\emph{location} to generically denote any of these three spatial aggregation 
levels. The creation of social connectivity maps between locations from widely 
accessible data, such as population geographic distribution (which is 
universally accessible for almost any region of the world through tools like 
Landscan), will prove useful for the study of information 
\cite{allen1984managing} or behaviour spreading in social networks among others 
\cite{fernandez2013voter}.

\subsection*{The effect of geography on social networks}

During the end of the 19th century and the beginnings of the
20th, a considerable amount of effort was dedicated to the development
of telecommunication systems. Such systems, whether they carried written
messages (telegraph) or voice (telephone), were designed to achieve a single
goal: allowing people to communicate with those who are far away (indeed
the Greek prefix \emph{tele-} means \emph{distant}). Interestingly  (and 
contrary to some predictions from the beginning of the internet era 
\cite{cairncross2001death} ) recent analyses of records from such system show 
that people do not commonly use them to talk to those far away, but with people 
who are actually close by.  Precisely, it has been consistently found across 
records from emails, phones and blogs that the probability of a communication 
to ocurr between two people who are $r$ kilometres apart from each other 
follows a decay function, typically a power law 
\cite{krings2009urban,herrera2013understanding,liben2005geographic,phithakkitnukoon2012socio}. 

Although the communication fluxes between regions have not been the focus of 
much research yet, the above mentioned new evidences show that communication 
fluxes behave similar to trip fluxes and other phenomena driven by the 
distribution of population across the geographical space. In transportation 
research, flux prediction is a well-defined problem: given a set of locations 
$\{i,j,...\}$ whose coordinates and populations $\{n_{i},n_{j},...\}$ are 
known, the goal is inferring the flux matrix $T$ where each element $T_{ij}$ 
represents the number of trips from
location $i$ to location $j$. The problem was traditionally approached
using gravity models \cite{erlander1990gravity,eggo2011spatial,balcan2009multiscale}
which try to gather the effect of decaying probability with distance $r_{i,j}$
following the equation \[
T_{ij}\propto\frac{n_{i}^{\alpha}n_{j}^{\beta}}{f(r_{ij})}
\]
where $\alpha$ and $\beta$ are fitting parameters usually estimated from 
training data, and $f(r_{ij})$ increases with distance, typically following an 
exponential or power-law function. A powerful idea was brought recently by the
radiation model \cite{simini2012universal}, which claims that it is not the
distance that matters, but the amount of opportunities between $i$
and $j$, which can be estimated by the population in the area. In
short, the authors explain that someone from rural Iowa is more likely to 
travel further to satisfy her needs than someone in New York City, given
the latter has a handful of options within a few blocks. While in its original 
publication the radiation model included testing against a phone call dataset 
(see Figure 3 in \cite{simini2012universal}), the problem of predicting 
communication fluxes was not the main focus of any model to date. In this paper 
we will present a new model inspired by this radiation model which is able to 
predict communication fluxes surprisingly well. Actually, the accuracy is 
similar to the one reached by current transportation models when predicting 
trips.

\section{Model description}

\begin{figure}
\begin{centering}
\includegraphics[width=0.5\columnwidth]{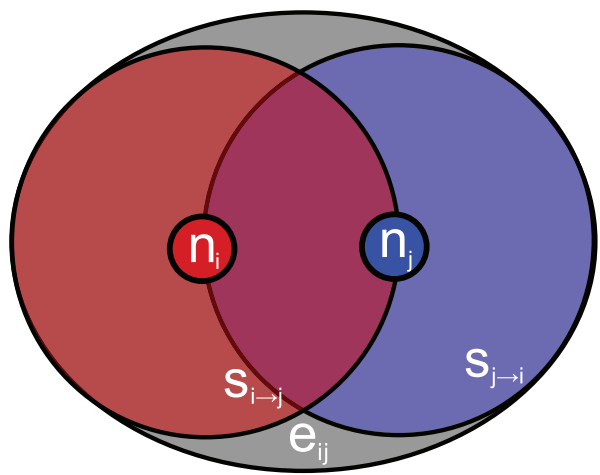}
\par\end{centering}

\caption{\label{fig:Model-scenario}Model scenario: $n_{i}$ represents the
population of location $i$ while $s_{i\rightarrow j}$ represents the
population within the circle with center in $i$ and radius up to
$j$. As long as population is not homogeneously distributed $s_{i\rightarrow j}\neq s_{j\rightarrow i}$, the radiation model predictions will not be symmetrical. $e_{ij}$
represents the population within the smallest ellipse whose focuses
are in $i$ and $j$ and contains both previous circles, as
well as $n_{i}$ and $n_{j}$.}
\end{figure}

Formally, the radiation model, when applied to social relationships, estimates 
the communication flux $T_{ij}$ between two locations $i$ and $j$ using the 
population in both locations, and the population within the circle whose center 
is $i$ and radius equal to the distance between $i$ and  $j$.  Its formulation
is 
\[
T_{i,j}^{rad}=K_{i}\frac{n_{i}n_{j}}{(n_{i}+s_{i\rightarrow j})(n_{i}+n_{j}+s_{i\rightarrow j})}
\]
where $n_{i}$ represents the population of location $i$, $s_{i\rightarrow j}$
the number of people who are not in $i$ but closer to $i$ than $j$
and the normalization $K_{i}=n_{i}\frac{N_{T}}{N}$, where $N_{T}$ is the total 
number of relationships to predict (which in general is considered to be 
available) and $N=\sum_{i}n_{i}$ the total population. 

It is straightforward to verify that $T^{rad}$ matrices are not symmetric in 
general.  While asymmetry is a desirable feature for mobility models (commuting 
origin-destination matrices have strongly asymmetric suburbs-downtown fluxes) 
it is not when dealing with communication fluxes, because the number of 
relationships people from location $i$ have with people from location $j$ must 
be the same as the number of relationships people from $j$ have with people 
from $i$. 

A natural modification of the radiation model to deal with communication fluxes 
could be a simple symmetrization of the model, which we will denote 
\emph{radBI} and whose formulation is
\[
T_{ij}^{radBI}=\frac{1}{2}(T_{ij}^{rad}+T_{ji}^{rad}).
\]

As shown below this model has a limited performance. This fact motivated us to 
develop the new model presented in this paper.
Our model, which we will refer as the elliptic model (EM), is oriented
to deal with social relationships. The EM considers that the probability of 
someone living at location \emph{i} having an acquaintance at location $j$ is 
reversely proportional to the population of the area where both could meet 
provided their combined travel distance does not exceed certain threshold. This 
area forms an ellipse whose focuses are in locations $i$ and $j$. Among all 
possible ellipses the model selects the smallest one containing the two 
$r_{ij}$ radius circles whose centers are in $i$ and $j$ respectively (see 
Figure \ref{fig:Model-scenario} for graphic explanation and comparison to the 
radiation model). Thus, the EM formulation is
\[
T_{ij}^{ellip}=K\frac{n_{i}n_{j}}{e_{ij}}
\]
where $e_{ij}$ is the population within the ellipse depicted in Figure
\ref{fig:Model-scenario} (note that $e_{ij}$ includes $n_{i}$ and $n_{j}$) and 
$K$ is a normalization parameter obtained from the total number of 
relationships to predict $N_{T}$ ( 
$\sum\limits_{i}\sum\limits_{j}T_{ij}^{ellip}=N_{T}$).  Since $e_{ij}=e_{ji}$, 
$T_{ij}^{ellip}=T_{ji}^{ellip}$ and thus our model produces symmetrical 
matrices $T$.

In order to compare the quantities involved in the model, one needs to consider 
the sets $S_{i \rightarrow j}$ and $S_{j \rightarrow i}$ such as $\#S_{i 
\rightarrow j}=s_{i \rightarrow j}$ and $\#S_{j \rightarrow i}=s_{j \rightarrow 
i}$. Lets address the case of a very large city $C\subset S_{i \rightarrow j}$ 
whose population $n_{C}\approx e_{ij}$.
While the radiation model  predicts different fluxes depending on whether $C 
\subset (S_{i \rightarrow j} \cap S_{j \rightarrow i})$ or not (smaller when 
$C$ belongs to the intersection) the EM will provide the same prediction for 
both cases. In fact, since $e_{ij} \geq \#(S_{i \rightarrow j} \cup S_{j 
\rightarrow i}) + n_{i} + n_{j}$ (and usually $e_{ij} \approx \#(S_{i 
\rightarrow j} \cup S_{j \rightarrow i}) + n_{i} + n_{j}$) the role of the 
union set is the main contribution of the model.

\section{Data description}

To evaluate the performance of the EM, we compare it with a mobile phone
data set containing Call Detail Records (CDRs) of a six month period in 3 
different countries:
France, Portugal, and Spain. In total, over 7 billion calls are considered
to build the social graph for each country, whose links are included only if 
there is at least one call in both directions during the observation period.  
The result is an undirected graph (this is a common technique in the literature 
to avoid both marketing callers and misdialed calls 
\cite{onnela2007structure}).  In Table \ref{tab:networks}, some characteristics 
of the networks are presented, such as high clustering and relatively low 
average degree, which are expected from previous literature about mobile phone 
networks.

\begin{table}

\begin{centering}
\begin{tabular}{lrrrrr}
\hline 
Country  & Users $U$  & Links $E$  & $\langle k\rangle$  & $\langle c\rangle$  & $\frac{N}{Total\, Population}(\%)$\tabularnewline
\hline 
France  & $18.7\cdot10^{6}$  & $81.3\cdot10^{6}$  & $8.73$ & $0.16$ & $30.21$\tabularnewline
Portugal  & $1.21\cdot10^{6}$  & $4.00\cdot10^{6}$  & $6.57$ & $0.26$ & $11.21$\tabularnewline
Spain  & $5.92\cdot10^{6}$  & $16.1\cdot10^{6}$  & $5.44$ & $0.21$ & $13.45$\tabularnewline
\hline \end{tabular}\caption{\label{tab:networks}Characteristic properties of 
the social networks in the studied countries: number of users (Nodes) and 
relationships (Links), average degree $\langle k\rangle$, average clustering 
coefficient $\langle c\rangle$ and relative sample size of the users in the 
data set.}

\par\end{centering}

\end{table}

In addition to the communication records, our data include a location
for each user: the most used mobile phone tower in France and Portugal
and the billing zip code in Spain. In order to benchmark the multi-scale 
performance of the EM, three aggregation levels have been used: country-wide 
fluxes between cities and regions and on the other hand metropolitan fluxes 
within cities.  Table \ref{tab:nloc} presents the number
of locations considered in each aggregation level. When applying these
spatial aggregations, the center of mass of the population is used
as the higher level location, instead of the centroid of the region
polygon (defined by administrative boundaries), in order to avoid undesirable 
effects in the fairly common case of a big city located in a corner of a 
polygon.

\begin{table}

\begin{centering}
\begin{tabular}{lrrr}
\hline Country  &  Towers/Zip codes & Cities  & Regions\tabularnewline
\hline 
France  & 17475 & 3520 & 96\tabularnewline
Portugal  & 2209 & 297 & 20\tabularnewline
Spain  & 8928 & 5446 & 52\tabularnewline
\hline \end{tabular}\caption{\label{tab:nloc}Number of locations considered in 
different geographic aggregation
levels for each country. At the finer level, mobile phone towers are available 
France and Portugal, and zip codes for Spain. Aggregation is based on 
administrative boundaries: cities are \emph{cántons }in France, \emph{concelhos 
}in Portugal and \emph{municipios }in Spain while regions mean 
\emph{départements} in France, and \emph{provincias }in Portugal and Spain. }

\par\end{centering}

\end{table}

\section{Communication fluxes in country scale}

To validate the predictions of the EM at large scale, we consider connectivity 
matrices $T$ in two aggregation levels. At the region level, matrix
$T$ has thousands of elements while at the city level there are tens
of millions of fluxes to predict (see Table \ref{tab:nloc}).
Input data for the predictions only consists of the location's coordinates
and populations, and the total number of relationships to predict $N_{T}$.  
Like the radiation model, the EM keeps the advantage of being parameter-free, 
so no training data is needed. 

\begin{figure}
\begin{centering}
\includegraphics[width=0.8\columnwidth]{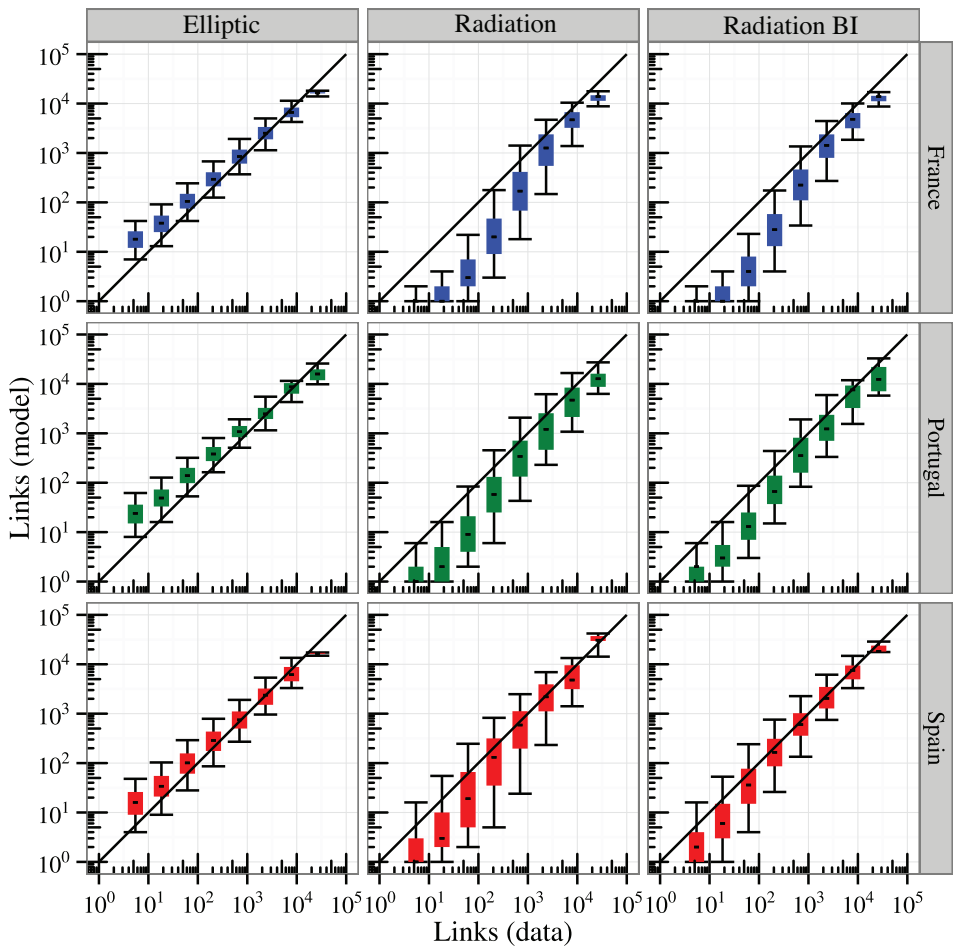}
\par\end{centering}

\caption{\label{fig:Predictions-by-different}Predictions by different models
versus real data. Fluxes between every city are presented, considering 297 
cities in Portugal\emph{, }5446 in Spain\emph{, }and
3520 in France\emph{. }Error bars plot 10\%, 30\%, 50\%, 70\% and
90\% quantiles. The elliptic model overcomes both radiation and bilateral
radiation models in all three scenarios. }
\end{figure}

In Figure \ref{fig:Predictions-by-different} we present a box-plot
of the predictions from all the three models versus real data for
fluxes between cities. The results prove consistently that the EM outperforms 
both the radiation model and its bilateral version. To present further evidence 
of the performance of the EM, we include in Table \ref{tab:country-res}
the $R^2$ of the predictions in both aggregation levels. The results
confirm that the EM outperforms previous models. 

Overall, the accuracy of the predictions is similar to the one obtained when 
applying transportation models to trip prediction 
\cite{simini2012universal,yang2013multiscale}. This is an unexpected finding, 
since in principle, while there is an increasing cost (like time or energy) 
associated with distance when travelling, there is not such cost when making a 
phone call. While, as stated in the introduction, there were previous reports 
illustrating that social ties depend on physical distance, the capability of 
reproducing a significant portion of the distribution of social ties between 
locations just by employing a map placing them and their populations, 
highlights even more the importance of the geographical space when forming 
ties. 

\begin{table}
\begin{centering}
\begin{tabular}{|c|r|r|r|r|r|r|}
\cline{2-7} 
\multicolumn{1}{c|}{} & \multicolumn{2}{c|}{France} & \multicolumn{2}{c|}{Portugal} & \multicolumn{2}{c|}{Spain}\tabularnewline
\cline{2-7} 
\multicolumn{1}{c|}{} & City & Province & City & Province & City & Province\tabularnewline
\hline 
Radiation & 0.534 & 0.615 & 0.621 & 0.776 & 0.556 & 0.588\tabularnewline
\hline 
RadiationBI & 0.626 & 0.723 & 0.730 & 0.847 & 0.676 & 0.668\tabularnewline
\hline 
Elliptic & 0.723 & 0.790 & 0.816 & 0.891 & 0.693 & 0.748\tabularnewline
\hline 
\end{tabular}
\par\end{centering}

\caption{\label{tab:country-res}$R^{2}$ of the different country-wide 
    predictions. Note that these $R^{2}$ are calculated without any logarithmic 
    transformation on data or predictions. The number of provinces considered 
    is 97, 20 and 52, respectively. Since the number of cities is up to two 
    orders of magnitude larger, the flux matrix $T$ is up to 4 orders of 
    magnitude larger.  While elliptic model is always more accurate than 
    previous models, the improvement is specially remarkable in fluxes between 
    cities.}
\end{table}

\section{Communication fluxes within cities}

While previous literature already stated that distance influences the creation 
of social ties between cities, our dataset allows us to study also urban 
relationships by using the finer spatial aggregation level available: phone 
towers or zip codes. Predicting all possible tower to tower relationships 
within the country would imply dealing with a $T$ matrix with up to 300 million 
elements, with only less than 1\% of them being not null. Thus, the prediction 
accuracy would be severely biased by the huge amount of zero cells. Instead, we 
study the short range accuracy of the model by applying it in each city where 
we got at least 20 different locations (the upper limit being Paris, where
we have over 1000 mobile phone towers). In total, the analysis includes 40 
cities in France, 29 Spain and 20 in Portugal.

Table \ref{tab:Average-for-urban} presents the results for the three algorithms 
in terms of average $R^{2}$. These results confirm that the EM still performs 
better, while the overall prediction accuracy is smaller compared to the 
country-wide experiment. The loss of accuracy within urban areas for any model 
purely based in distance is expected and observed in the transportation field 
\cite{yang2013multiscale}. One of the main reasons for this loss of accuracy is 
the fact that the distance is a poorer proxy for travel time or cost in cities.  
People in cities tend to be within a daily radius of action and the decision of 
whom they communicate with depends on other metrics related to the different 
hierarchies that could define a social distance (e.g. ethnicity, occupation, 
etc.) \cite{herrera2013understanding}.
\begin{table}
\begin{centering}
\begin{tabular}{|c|r|c|c|}
\cline{2-4} 
\multicolumn{1}{c|}{} & \multicolumn{1}{c|}{France} & \multicolumn{1}{c|}{Portugal} & \multicolumn{1}{c|}{Spain}\tabularnewline
\hline 
Radiation & 0.377 & 0.527 & 0.434\tabularnewline
\hline 
RadiationBI & 0.436 & 0.608 & 0.498\tabularnewline
\hline 
Elliptic & 0.653 & 0.658  & 0.501\tabularnewline
\hline 
\end{tabular}
\par\end{centering}

\caption{\label{tab:Average-for-urban}Average $R^{2}$ for urban fluxes prediction
for every city in the data set where there are at least 20 different
locations (towers or zip codes). The number of locations range from this
20 up to 1000 in Paris. This sums up to 40 cities in France, 29 Spain
and 20 in Portugal. Although the EM again outperforms previous models,
each performance is small when compared to country-wide scenarios. }
\end{table}

\subsection*{Correction $\varepsilon$ as a model improvement for urban areas}

\begin{figure}
\begin{centering}
\includegraphics[width=1\columnwidth]{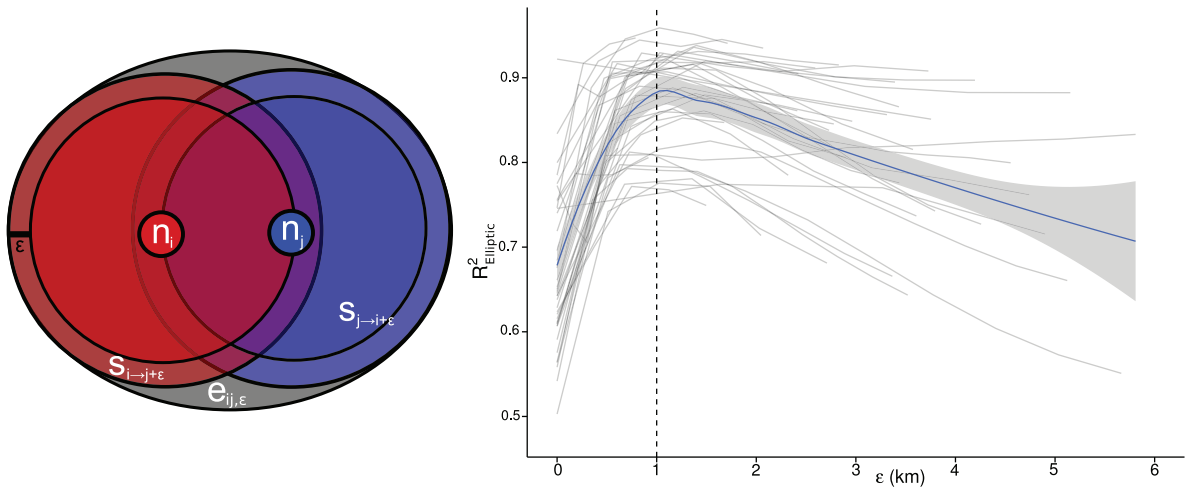}
\par\end{centering}

\caption{\label{fig:Model-scenario-1}Model modified for intracity predictions,
adding the correction term $\varepsilon$. Each grey line represents a certain 
city in the dataset with the blue line and the shadow representing the general 
trend.  We find predictions improve when some correction term is included, 
reaching a maximum around $\varepsilon=1\mbox{km}.$ }
\end{figure}

When applying the EM based on Figure \ref{fig:Model-scenario} to urban 
relationships one should be aware that a tower $k$ whose distance to tower $i$ 
is $r_{ik}=r_{ij}+\varepsilon$ where
$\varepsilon\ll r_{ij}$ will not be taken into account for predicting $T_{ij}$.  
Since towers tend to be closer to each other in urban areas, we propose the 
correction in Figure \ref{fig:Model-scenario-1} for urban environments. The 
variation consists of including a correction parameter $\varepsilon$ so that 
the ellipse is now the smallest one containing the two circles of radius 
$r_{ij} + \varepsilon$ centered in $i$ and $j$.
After studying several values of $\varepsilon$, we found that the prediction 
accuracy peaks near $\varepsilon=1$km for nearly all the cities (as shown in 
Figure \ref{fig:Model-scenario-1}).  

There may be several interpretations for such a maximum: one could argue that 
it comes from the location error, known to be close to the average distance to 
neighbors from the Voronoi tessellation \cite{ulm2013properties}, which is 
around 1
km in average for our dataset. This agrees with the fact that the optimal 
$\varepsilon$ is a fixed value and does not depend on the distance $r$ between 
$i$ and $j$.  On the other hand, when applied back to country-wide scenarios we 
found the correction term does not improve the predictions and no peak emerges 
near $\varepsilon\backsimeq r_{Voronoi}$ or elsewhere, reinforcing the 
hypothesis that within cities we are reaching the boundaries of the resolution 
of our location data.

\begin{figure}
\begin{centering}
\includegraphics[width=1\columnwidth]{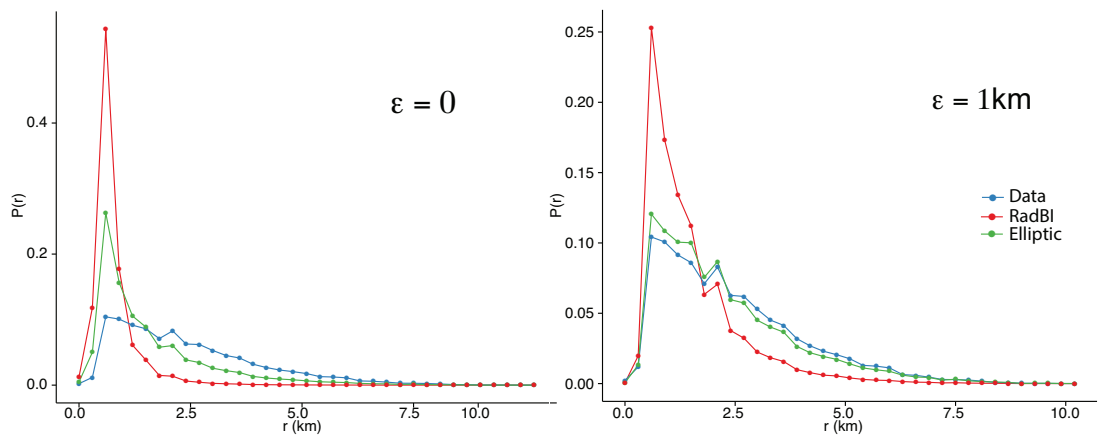}
\par\end{centering}

\caption{\label{fig:pr}Left: fraction of relationships $P(r)$ within distance 
    $r$ in the real dataset compared to predictions
by both elliptic and bilateral radiation models where $\varepsilon=0$
for Porto (Portugal). Right: elliptic prediction gets very close to data when 
using $\varepsilon=1\mbox{km}$. Although the radiation model predictions also 
improve, they still predict shorter relationships than those observed in 
reality.  }
\end{figure}

Another way to evaluate the performance of the different models is to compare 
them against empirical data in terms of the link-distance distribution $P(r)$ 
which represents the probability of observing a relationship between two people 
living $r$ kilometres from each other. Figure \ref{fig:pr} shows the 
improvement  $P(r)$ when applying the $\varepsilon=$1km correction. Without the 
correction term, short-range relationships are over represented, while the EM 
with the correction fits almost perfectly with the distribution obtained from 
the data.  Note that although radiation model predictions also improve, it 
still predicts shorter relationships than those observed in the data. 

Table \ref{tab:epsilon} shows results of the corrected model for urban 
environments in terms of average $R^{2}$, which confirm a significant 
performance increase when applying the corrected model with $\varepsilon=1$ km 
across all cities in the data set. 

\begin{table}
\begin{centering}
\begin{tabular}{|c|r|c|c|}
\cline{2-4} 
\multicolumn{1}{c|}{} & \multicolumn{1}{c|}{France} & \multicolumn{1}{c|}{Portugal} & \multicolumn{1}{c|}{Spain}\tabularnewline
\hline 
Elliptic $\varepsilon=0$ & 0.670 & 0.645  & 0.494\tabularnewline
\hline 
Elliptic $\varepsilon=1\,\mbox{km}$ & 0.846 & 0.740 & 0.688\tabularnewline
\hline 
\end{tabular}
\par\end{centering}

\caption{\label{tab:epsilon} Average $R^{2}$ of the predictions for the 
    corrected model with $\varepsilon=1\mbox{km}$, compared to the original 
    (non-corrected) model.}
\end{table}

\section{Conclusions and further research}
In this paper the problem of predicting communication fluxes between different 
locations has been successfully addressed. A new model has been proposed to 
calculate the communication fluxes using only population distribution data.  
This data is publicly available worldwide through projects which provide 
population estimates for almost every square mile on earth.

The presented model successfully takes into account the symmetry of the 
communication fluxes, in order to predict the number of social ties between 
geographical locations at different scales, ranging from neighborhoods to 
regions. Interestingly, we have shown that geolocated population data is as 
useful to predict communication fluxes as it is to predict trip fluxes.  

The proposed model is readily available to be used by researchers in different 
social sciences studying various phenomena where human ties are known to be 
crucial such as information propagation or disease spreading. Overall, our 
model implies social ties are to a large extent driven by geographical factors.  
While there may be other factors influencing very long distance relationships 
(e.g.  time zones, or natural, national \cite{ugander2011anatomy} and language 
borders \cite{blondel2010regions}, etc.) the available data did not allow to 
check them, so that further research would be needed along this line.

In order to enhance the employment of the EM by the research community, 
implementations in three widely used programming languages have been made 
available on our homepage \cite{ellipticwell}, together with an interactive 
tool with France \emph{départments} as an example scenario.

\section{References}

\bibliographystyle{unsrt}
\bibliography{references}

\end{document}